\documentclass[useAMS,usenatbib,usegraphicx]{mnras}
\usepackage{epsfig}
\usepackage{float}
\usepackage{placeins}
\usepackage{graphicx}
\usepackage{epsfig}
\usepackage{bm}
\usepackage{amsfonts}
\usepackage{amssymb}
\usepackage{times}
\usepackage{natbib}
\usepackage{color}
\usepackage{amsmath}
\usepackage{verbatim}

\title[Multi-wavelength flare in 3C 454.3]{A peculiar multi-wavelength flare in the Blazar 3C 454.3}
\author[Gupta et al.]
{Alok C.\ Gupta$^{1,2}$\thanks{E-mail: acgupta30@gmail.com}\thanks{CAS PIFI Visiting Scientist},
Arun Mangalam$^{3}$\thanks{E-mail: mangalam@iiap.res.in},
Paul J.\ Wiita$^{4,5}$\thanks{E-mail: wiitap@tcnj.edu}, 
P.\ Kushwaha$^{6}$, H.\ Gaur$^{1}$,
\newauthor H.\ Zhang$^{7}$, M.~F.\ Gu$^{1}$, M. Liao$^{1,8}$, G.\ Dewangan$^{9}$,
L. C.\ Ho$^{10,11}$, P.\ Mohan$^{12}$, M.\ Umeura$^{13}$, 
\newauthor M.\ Sasada$^{14}$, A. E.\ Volvach$^{15,16}$,
A.\ Agarwal$^{2}$, M. F.\ Aller$^{17}$, H.~D.\ Aller$^{17}$, R.\ Bachev$^{18}$, 
\newauthor
A.\ L{\"a}hteenm{\"a}ki$^{19}$, E.\ Semkov$^{18}$, A.\ Strigachev$^{18}$, M.\ Tornikoski$^{19}$,
L.~N.\ Volvach$^{15,16}$
\\
$^{1}$Key Laboratory for Research in Galaxies and Cosmology, Shanghai Astronomical Observatory, Chinese Academy
of Sciences, Shanghai 200030, China \\
$^{2}$Aryabhatta Research Institute of Observational Sciences (ARIES), Manora Peak, Nainital -- 263002, India \\
$^{3}$Indian Institute of Astrophysics, Sarjapur Road, Koramangala, Bangalore -- 560034, India \\
$^{4}$Department of Physics, The College of New Jersey, P.O.\ Box 7718, Ewing, NJ 08628-0718, USA \\
$^{5}$Kavli Institute for Particle Astrophysics and Cosmology, SLAC, Menlo Park, CA 94025, USA \\
$^{6}$Department of Astronomy (IAG-USP), University of S{\~a}o Paulo, S{\~a}o Paulo 05508-900, Brazil \\ 
$^{7}$Astrophysical Institute, Department of Physics and Astronomy, Ohio University, Athens, OH 45701, USA \\
$^{8}$University of Chinese Academy of Science, 19A Yuquanlu, Beijing 100049, China \\
$^{9}$Inter University Centre for Astronomy and Astrophysics (IUCAA), Post Bag 4, Ganeshkhind, Pune -- 411007, India \\
$^{10}$Kavli Institute for Astronomy and Astrophysics, Peking University, Yi He Yuan Lu 5, Hai Dian District, 
Beijing 100871, China \\ 
$^{11}$Department of Astronomy, Peking University, Yi He Yuan Lu 5, Hai Dian District, Beijing 100871, China \\
$^{12}$Shanghai Astronomical Observatory, Chinese Academy of Sciences, Shanghai 200030, China \\
$^{13}$Hiroshima Astrophysical Science Center, Hiroshima University, Kagamiyama 1-3-1, Higashi-Hiroshima 739-8526, Japan \\
$^{14}$Department of Astronomy, Graduate School of Science, Kyoto University, Kitashirakawa-Oiwake-cho, Sakyo-ku, 
Kyoto 606-8502, Japan \\
$^{15}$Radio Astronomy Laboratory of Crimean Astrophysical Observatory, Nauchny, Crimea \\
$^{16}$Taras Shevchenko National University of Kyiv, 4, Academician Glushkov Ave., 03127, Kiev, Ukraine \\
$^{17}$Astronomy Department, University of Michigan, 311 West Hall, Ann Arbor, MI 48109-1107, USA \\
$^{18}$Institute of Astronomy and National Astronomical Observatory, Bulgarian Academy of Sciences, 
72 Tsarigradsko Shosse Blvd., 1784 Sofia, Bulgaria \\
$^{19}$Aalto University, Mets{\"a}hovi Radio Observatory, P.O.\ Box 11000, FI-00076, Aalto, Finland}

\begin{document}
\newdimen\digitwidth
\setbox0=\hbox{2}
\digitwidth=\wd0
\catcode `#=\active
\def#{\kern\digitwidth}

\date{Accepted 2017 August 08. Received  2017 August 08; in original form 2017 April 20}

\pagerange{\pageref{firstpage}--\pageref{lastpage}} \pubyear{2017}

\maketitle

\label{firstpage}

\begin{abstract}

The blazar 3C454.3 exhibited a strong flare seen in $\gamma$-rays, X-rays, and
optical/NIR bands during 3--12 December 2009. Emission in the V and J bands
rose more gradually than did the  $\gamma$-rays and soft X-rays, though all peaked
at nearly the same time. Optical polarization measurements showed dramatic changes
during the flare, with a strong anti-correlation between optical flux and degree
of polarization  (which rose from $\sim$3\% to $\sim$ 20\%) during the declining
phase of the flare.  The flare was accompanied by  large rapid swings in polarization
angle of $\sim$ 170$^{\circ}$. This combination of behaviors appear to be unique. 
We have cm-band radio data during the same period but they show no correlation with  
variations at higher frequencies. Such peculiar behavior may be explained using 
jet models incorporating fully relativistic effects with a dominant source region 
moving along a helical path or by a shock-in-jet model incorporating three-dimensional 
radiation transfer if there is a dominant helical magnetic field. We find that spectral 
energy distributions at different times during the flare can be fit using modified one-zone 
models where only the magnetic field strength and particle break frequencies and 
normalizations need change. An optical spectrum taken at nearly the same time provides 
an estimate for the central black hole mass of $\sim$ 2.3 $\times$ 10$^{9}$M$_{\odot}$.
We also consider two weaker flares seen during the $\sim 200$ d span over
which multi-band data are available.  In one of them, the V and J bands appear to
lead the $\gamma$-ray and X-ray bands by a few days;  in the other, all variations
are simultaneous.
 
\end{abstract}


\section{Introduction}
The flat spectrum radio quasar (FSRQ) 3C 454.3 ($2251+158; z = 0.859$) is a bright
and frequently observed blazar.  It shares the common FSRQ properties of non-thermal
emission and significant variability across the entire electromagnetic spectrum
along with substantial optical polarization (Smith et al.\ 1988; Healey et al.\
2007; Sasada et al. 2013, 2014).  FSRQs such as 3C 454.3 have spectral energy distributions (SEDs) that
show the usual two broad humps
peaking at mm-IR wavelengths and around 1 GeV (Urry \& Padovani 1995; Sambruna
et al.\ 1996). The lower energy one is ascribed to synchrotron emission from a
relativistic jet pointing near our line of sight, while the high energy peak presumably
arises from inverse Compton (IC) scattering of lower energy photons off the synchrotron
emitting relativistic particles (e.g., Urry \& Padovani 1995). The strongest constraints
on emission models and the locations of emission regions in blazars can come from
 analysis of broadband SEDs as they vary in time, and any clear
correlations between different bands during flares are of particular interest. The
central engine of 3C 454.3 contains a super-massive black hole estimated in the
range $0.5 - 1.5 \times 10^9$ M$_{\odot}$ (Woo \& Urry 2002; Liu et al.\ 2006;
Sbarrato et al.\ 2012).  The flow speed down the approaching relativistic jet is
probably between $0.97c$ and $0.999c$ (Jorstad et al.\ 2005; Hovatta et al.\ 2009; Raiteri et al.\  2011)
and the angle to our line of sight is between $1^\circ$ and  $6^\circ$ (Raiteri et al.\  2011; Zamaninasab et al.\ 2013).

Several earlier multi-band observations of 3C 454.3 have been conducted. Those
including simultaneous $\gamma$-ray fluxes are of interest to us here, and as they must
incorporate Fermi or AGILE data they must be relatively recent.  In the observations
of Bonning et al.\ (2009) excellent correlations between IR, optical, UV and
$\gamma$-ray fluxes were seen, with lags within one day; however, the X-ray flux then
was almost non-variable and not correlated with either the higher or lower frequency
measurements. Vercellone et al.\ (2009) also saw correlated optical and high energy
$\gamma$-rays measured by AGILE; they had INTEGRAL and Swift X-ray measurements,
though the latter were again not well correlated.  More complete AGILE-led multi-band
monitoring of 3C 454.3 over 20 months  (Vercellone et al.\ 2010; Raiteri et al.\ 2011)
found nearly
simultaneous flux peaks across all bands from mm to $\gamma$-rays during the strong flares, with the
$\gamma$--optical correlation usually having a time-lag less than a day. Strong correlations
between $\gamma$-ray and optical light curves (LCs) were found by Gaur et al.\ (2012),
though in that case, the $\gamma$-ray LC led the optical one by $4.5 \pm 1.0$ days.
Again, the X-ray LC was essentially constant and so showed no correlation with
the other bands. Similar strong correlations were found between NIR-optical
and $\gamma$-rays by Kushwaha et al.\ (2017), but with Fermi-LAT $\gamma$-rays lagging
the optical-NIR by $\sim 3$ days. While in this case the X-rays showed a behavior similar to that 
observed in the optical-NIR,
they were not well sampled. Strong correlated flux variability between
Fermi $\gamma$-rays and 37 GHz radio flares have been seen on different occasions (Le{\'o}n-Tavares 
et al.\ 2011; Ramakrishnan et al.\ 2015) in this blazar.  The above studies of 3C
454.3 did not include optical polarization measurements.
 
A detailed multi-band analysis of the variability of 3C 454.3 between
2009 and 2011 that did include some optical spectropolarimetry (Jorstad et al.\ 2013)
discovered similar triple flare structures for each of three $\gamma$-ray outbursts.
These correlations indicate that the locations and mechanisms are similar for all of
those flares, the first one of which in December 2009 we revisit here with the incorporation of substantial
additional optical photometry and polarimetry.  Radio knots in the inner jet were
associated with the first and third outbursts in mm-bands (Jorstad et al.\ 2013)
and here we also include cm-band light curves.  Other studies of this large flare of
3C 454.3 have included X-ray data from Swift-BAT, INTEGRAL-IBIS and HEXT and 
optical data from Swift-UVOT (Pacciani et al.\ 2010).  Sasada et al.\ (2012) also studied
this flare with a focus on optical polarimetric variations.  

We now briefly note several multi-band measurements of other blazars that incorporated
optical polarimetry measurements.  Another bright FSRQ that showed a $\gamma$-ray flare in 
Fermi observations during a multi-band campaign is 3C 279 (1253$-$055; $z = 0.536$); Abdo et 
al.\ (2010) noted that it was coincident with a large change of optical polarization.  In this 
instance the $\gamma$-ray
flux peaked shortly before the optical and NIR fluxes and once again there were no
significant simultaneous X-ray or radio variations at that time; however, there
was a strong X-ray flare some two months later that might have had very modest
optical/NIR and $\gamma$-ray counterparts. On another occasion in 2011 3C 279 was in a high 
$\gamma-$ray activity state that showed multiple peaks and coincided exactly with a 352$^{\circ}$ rotation of 
the optical polarization angle and flaring activity at optical bands (Kiehlmann et al.\ 2016).
The prototype BL Lacertae object, BL Lac, has of course also been subject to a great deal of 
multi-band monitoring.
Marscher et al.\ (2008) made multiple VLBI radio maps and optical polarization
measurements and were able to detect a knot in the jet that apparently produced
a double flare that emitted between optical and TeV $\gamma$-ray energies along
with a radio outburst detected later.   Another peculiar result for BL Lac was the
discovery of a phase in its optical LC where the flux strongly anti-correlated with
the degree of optical polarization (PD) while the angle of polarization stayed
essentially fixed (Gaur et al.\ 2014). Finally, we mention  multi-wavelength 
($\gamma$-ray, optical and optical polarization, plus VLBA) variations of another
bright BL Lac, S5 0716$+$714 (Larionov et al.\ 2013).  They found rapid rotation of
the linear polarization vector to coincide with a peak in both $\gamma$-ray and
optical fluxes and that a new superluminal radio knot appeared at essentially the
same time (see also Chandra et al.\ 2015).

In Section 2 of this paper we bring together $\gamma$-ray, X-ray, optical/NIR
and radio flux measurements for 3C 454.3, along with optical polarimetry, during the
period $\sim$ MJD 55000 -- MJD 55200.   During one substantial flare an apparently
unique combination of flux and polarization changes were detected.  In Section 3
we discuss models that could produce such observations.

\section{Data and Results}

\subsection {Gamma-ray fluxes}
The Large Area Telescope (LAT) on the Fermi Gamma Ray Space Telescope (hereafter Fermi-LAT;
Atwood et al.\ 2009) has been observing the gamma-ray sky since its launch in June
2008. The high sensitivity and wide field of view ($\sim 2.4$ steradians) of Fermi-LAT
means that it has revolutionized our knowledge of the sky in its energy band that covers
20 MeV to 300 GeV. Fermi-LAT normally operates in a scanning mode that covers the entire
sky every three hours.

The gamma-ray light curve of 3C 454.3 is shown in the top panel of Figure 1. We extracted
daily gamma-ray fluxes in the 100 MeV to 300 GeV energy range using the
standard unbinned likelihood method implemented in the pylikelihood library of the
Fermi science tools version 10r0p5. Our analysis considers only the SOURCE class
events tagged as ``evclass=128, evtype=3'' under the PASS 8 instrument response
function P8R2\_SOURCE\_V6 from a circular region of interest (ROI) $15^\circ$ centered on
the source location. The Earth's limb $\gamma$-ray background was minimized by avoiding
photons arriving from zenith angle of $> 90^\circ$ while satellite operation and data
acquisition quality was insured using filter ``(DATA\_QUAL$>$0)\&\&(LAT\_CONFIG==1)''. The
source model file was generated from the 3rd LAT catalog (3FGL --gll\_psc\_v16.fit; Acero et al. 2015) 
incorporating the Galactic
and isotropic extragalactic $\gamma$-ray background by the respective templates
\emph{gll\_iem\_v06.fits} and \emph{iso\_P8R2\_SOURCE\_V6\_v06.txt} provided by the
LAT team. The effect of other sources outside the ROI was accounted by
generating an exposure on an additional annulus of 10 degrees around it. With all these
inputs the likelihood fit was iteratively performed by removing point sources that were
not contributing at the time and thus had a test statistic
(TS; Mattox et al.\ 1996) $< 0$ (Kushwaha et al.\ 2014).  
We used a log-parabola
model for the source with all parameters being free and the converged best fit was
used to derive the photon flux.

\subsection {X-ray fluxes}

Swift XRT data were gathered in the pointed photon counting (PC) and  windowed timing
(WT) modes. Most of the PC data have rates requiring pile up correction while many
WT data show varied position angles within one observation. We, thus, used the XRT
data files generated from the online tools described by Evans et al.\ (2009) 
available at the UK Swift Science Data Center\footnote{\tt http://www.swift.ac.uk/user$\_$objects}. It corrects
the products for bad pixels and columns, pile up and field of view effects. Additionally,
for finding the spectrum, this reduction also provides count-weighted auxiliary response and response matrix
files accounting for the off axis angle effects. The pile-up corrections for the PC data
are done by using an annular source extraction region of varying inner radius depending
on the rate. The resulting spectrum file for each observation was then modeled with the
\emph{phabs*power-law} model in \emph{XSPEC} (version 12.9.0i) with a fixed neutral
hydrogen column density of $1.34\times 10^{21} ~\rm cm^{-2}$ (Villata et al. 2006). 
The unabsorbed flux between $0.3-10$ keV was then calculated using the \emph{cflux}
task. The resultant light curve is shown in the second panel of Fig.\ 1.

\subsection{Optical/NIR photometry }

We  obtained photometric
observations of 3C 454.3  from the TRISPEC instrument mounted on the 1.5 m ``KANATA''
telescope at Higashi-Hiroshima Observatory. TRISPEC is able to perform simultaneous
three-band (one optical and two NIR bands) imaging or spectroscopy along with
polarimetry (Watanabe et al.\ 2005; Sasada 2012). The V and J band photometric data 
from the KANATA telescope are presented in red and cyan symbols, respectively, in the fifth
panel of Fig.\ 1.

We also used the publicly available 
SMARTS\footnote{\tt http://www.astro.yale.edu/smarts/glast/tables/3C454.tab}
data where the observations are carried out on the 1.3m telescope located at Cerro
Tololo Inter-American Observatory with the ANDICAM instrument. Data reduction and
analysis of SMARTS data  is described in Bonning et al.\ (2012). The V and J band
SMARTS data are represented by the green and magenta symbols, respectively. 

Finally, we also include optical V band public archival observations from Steward 
observatory\footnote{\tt http://james.as.arizona.edu/$\sim$psmith/Fermi/DATA/Objects/} 
(Smith et al.\ 2009); these are represented by black symbols in the fifth panel of Fig.\ 1.  
As can be seen in Fig.\ 1 there is good agreement between measurements made at the different
telescopes during times when more than one of these observatories obtained data.

\subsection{ Optical polarimetry }

We obtained polarimetric observations of 3C 454.3 at the 1.5 m ``KANATA" telescope (Sasada et al.\ 2012).
The polarization parameters are calculated from four consecutive images, which were
obtained with half-wave-plate angles of 0$^{\circ}$, 22.5$^{\circ}$, 45$^{\circ}$
and 67.5$^{\circ}$. The instrumental polarization was less than 0.1\% in the V band.
These data are presented in red symbols in the third (polarization angle, PA) and fourth
(polarization degree, PD) panels of Fig.\ 1.  

Sparser data from the Steward Observatory spectropolarimetric monitoring project
(Smith et al.\ 2009) were previously published (Jorstad et al.\ 2013) and are also
shown here (with black symbols) in the third and fourth panels of Fig.\ 1. There is 
excellent agreement between the polarimetric measurements during the limited periods of
overlap.

\subsection{Radio observations}

The 22.2 GHz radio observations of 3C 454.3 were carried out with the 22-m radio telescope (RT-22)
of the Crimea Astrophysical Observatory (CrAO) (Volvach 2006). Modulated radiometers were 
used in combination with the ``ON-ON'' registration regime (Nesterov, Volvach \& Strepka 2000). 
Radio observations at 36.8 GHz were made with the 14-m radio telescope (RT-14) of Aalto University 
Mets{\"a}hovi Radio Observatory in Finland. 
A detailed description of the data reduction and analysis of Mets{\"a}hovi data is given in Teraesranta 
et al.\ (1998). The observations and data processing techniques are similar for both RT-22 and RT-14.   
The results are given in the bottom panel of Fig.\ 1 in blue symbols for 22.2 GHz and magenta symbols 
for 36.8 GHz frequency.  

Radio observations at 14.5, 8.0 and 4.8 GHz were obtained from the University of Michigan Radio 
Astronomical Observatory (UMRAO) (Aller et al.\ 1999, 2014) which provided well sampled radio LCs 
at those frequencies for $\sim$100 AGNs over time spans of $\sim$30 yrs.

\subsection{Combined Multi-wavelength Results}\label{subsec:results}

\begin{figure*}
\centering
\includegraphics[width=180mm]{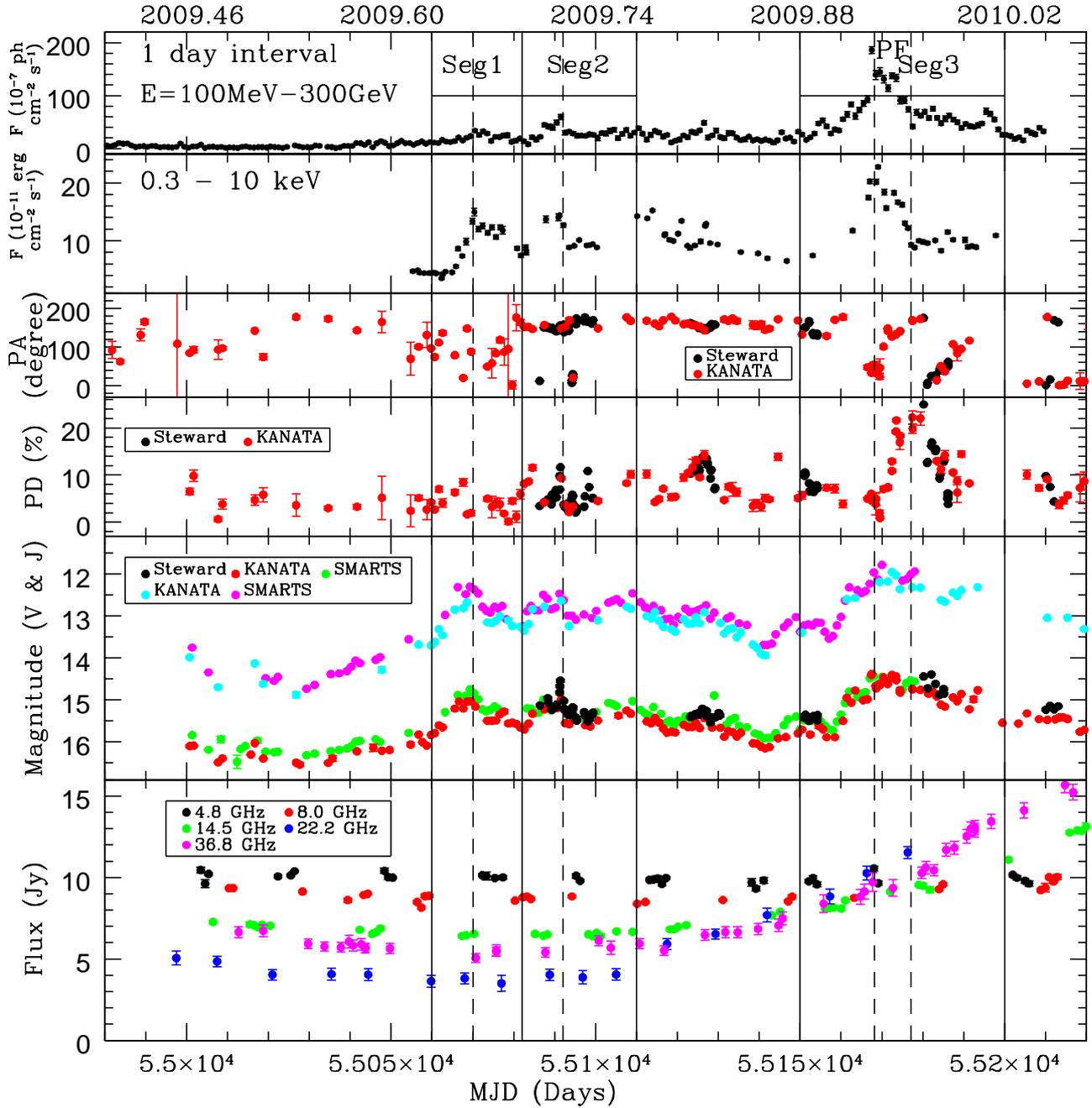}
\caption{In the top and second panels  $\gamma-$ray and X-ray fluxes from  Fermi-LAT  and Swift/XRT
are respectively presented.  The third panel gives optical position angle (PA) while the
fourth gives the optical polarization degree (PD). In the fifth panel, we give optical V band
data from KANATA, SMARTS and Steward as the lower light curve along with near-infrared J band data from KANATA and SMARTS
as the upper one.
In the bottom panel radio data are given from: UMRAO at 4.8 GHz, 8.0 GHz, and
14.5 GHz;  CrAO at 22.2 GHz; and  Mets{\"a}hovi at 36.8 GHz.  Except where shown, error bars are 
smaller than the symbols.  Solid vertical lines mark the three segments in which multi-band correlations
are studied and dashed vertical lines respectively mark the peaks of the first and second flares, the peak of the 
peculiar flare (PF), and a decaying phase of PF.}
\label{data}
\end{figure*}

\begin{figure*}
\centering
 \includegraphics[width=75mm,height=95mm]{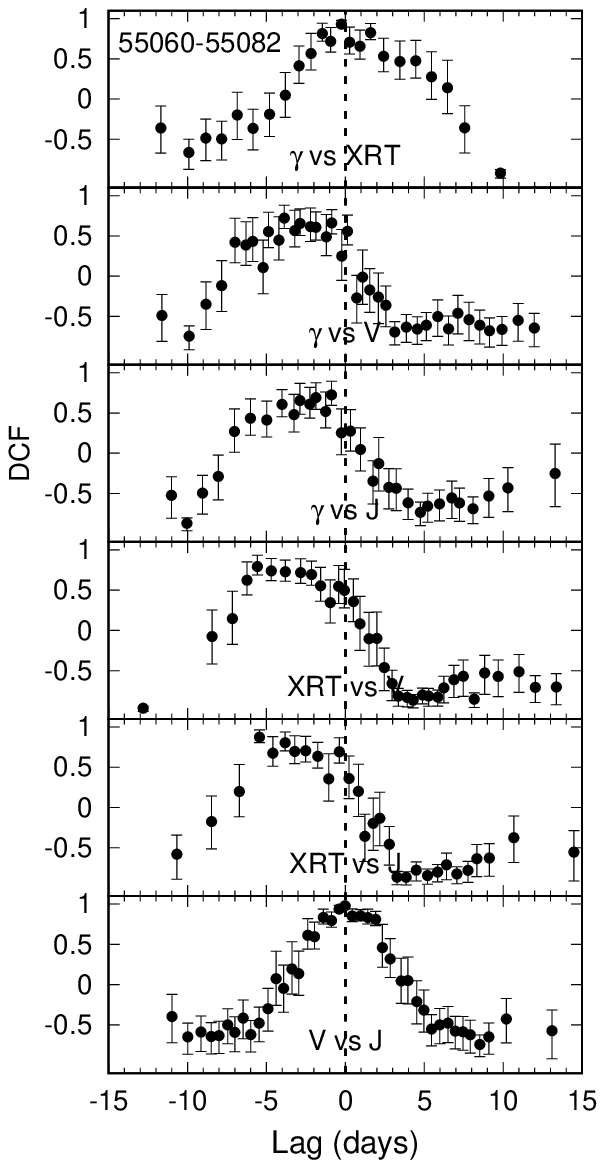}
 \includegraphics[width=75mm,height=95mm]{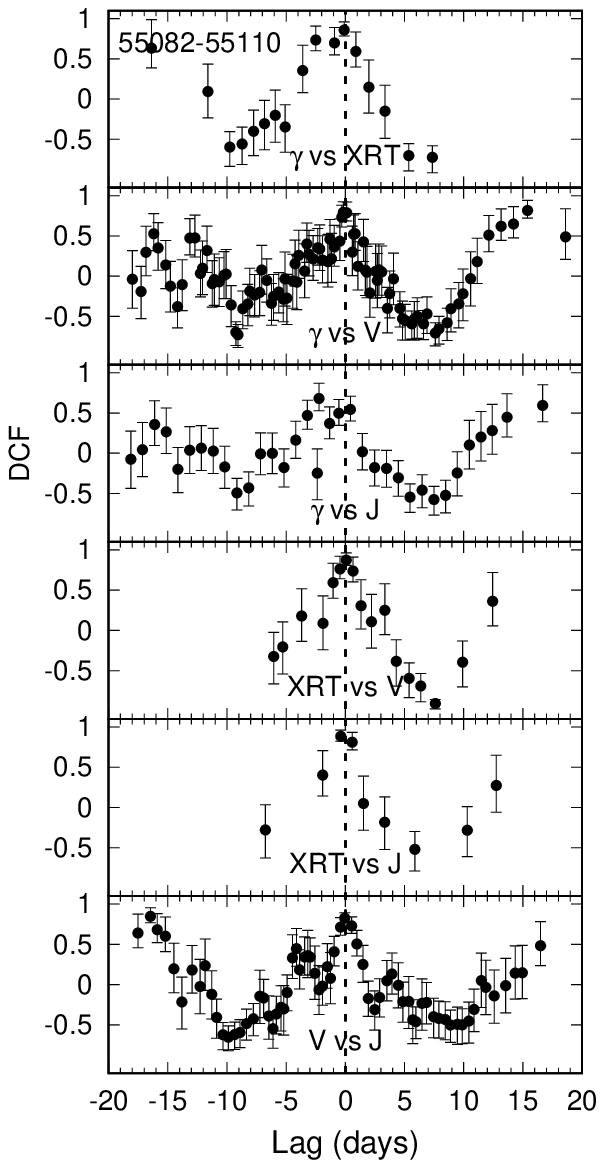}
 \includegraphics[width=75mm,height=95mm]{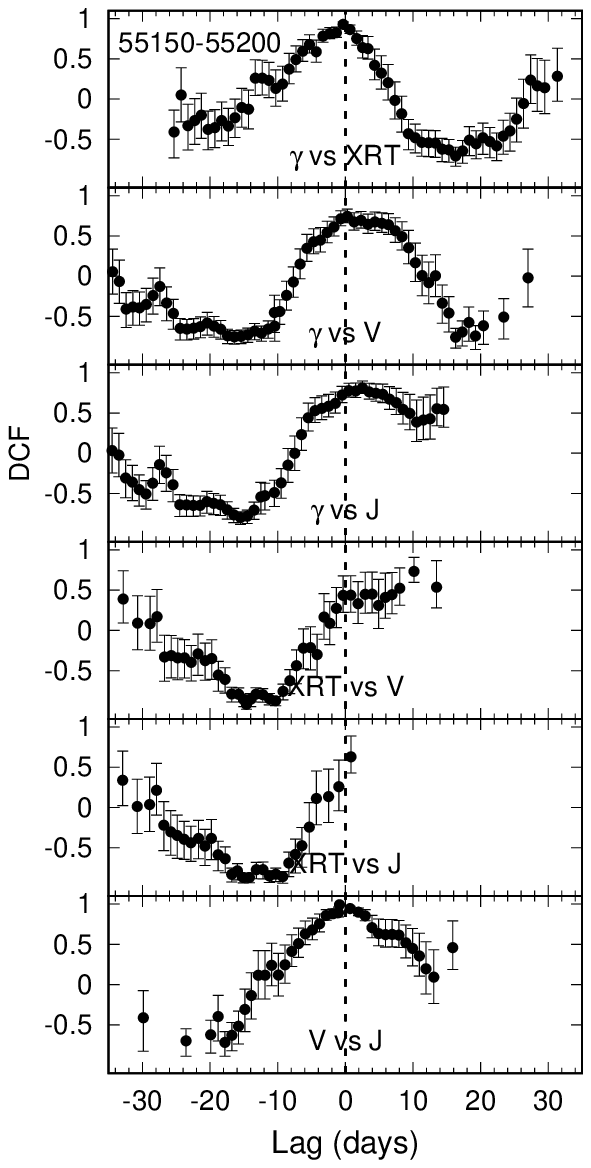}
 \includegraphics[width=75mm,height=95mm]{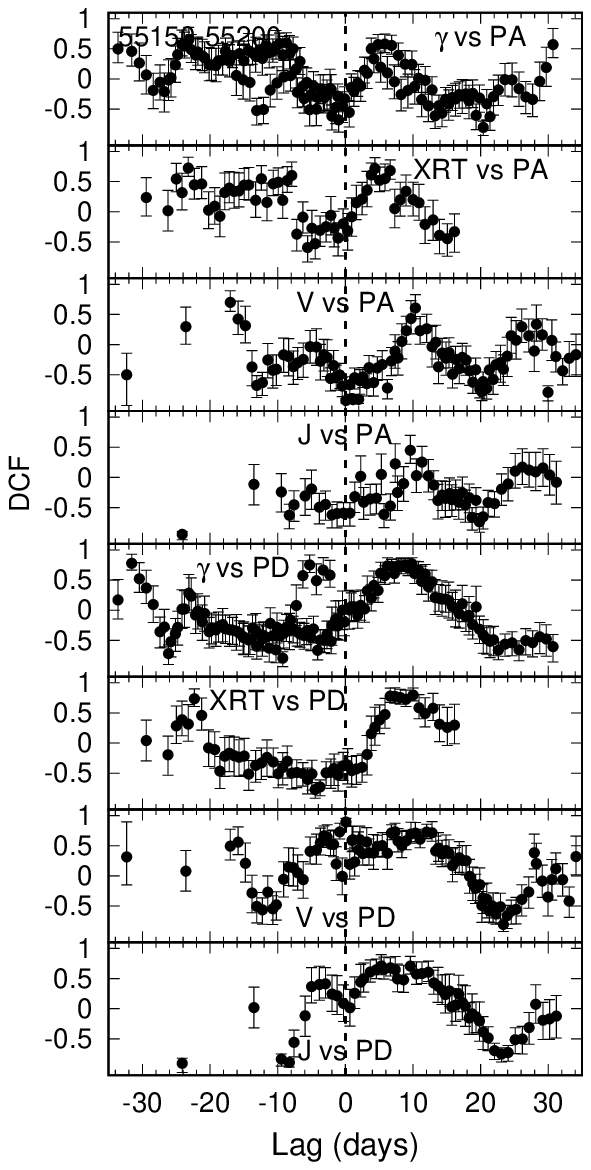}
 \caption{DCFs for the 3 segments of the multi-wavelength light curves of 3C 454.3
 (see Figure \ref{data}) between $\gamma$-ray vs optical (V), X-ray and NIR (J);
 X-ray vs optical (V) and NIR (J), and optical (V) vs NIR (J). The last panel shows
 the DCFs between light curves with PA and PD for the segement 3. The vertical dashed
 line marks the zero lag between the two light curves (see \S\ref{subsec:results}).}
 \label{fig:zdcf}
\end{figure*}

We show the multi-wavelength ($\gamma$-ray, X-ray, optical, NIR, radio fluxes and optical
polarization data) taken during MJD 54980 -- 55220 in Fig.\ 1. On visual inspection, three flares
are seen and we focused
our study on them. To search for cross-correlated variability, we selected three segments 
from the whole data presented in Fig.\ 1. Searching for cross-correlated variability, the discrete 
correlation functions (DCFs) were estimated using the \emph{z-transformed discrete correlation 
function} \citep[ZDCF;][]{1997ASSL..218..163A, 2013arXiv1302.1508A} method, applicable to both 
uniformly and non-uniformly sampled data. The results for these  three selected
segments are reported in  Table \ref{tab:lagResults} and Figure \ref{fig:zdcf}.

\subsubsection{Segment 1 (MJD 55060 -- 55082)}    

In Fig. 1, it is clear that there is a  strong X-ray flare starting at $\sim$ MJD 55060 and peaking at 
$\sim$ MJD 55070 which then declines and reached a  minimum flux state at $\sim$ MJD 55082. 
To search for multi-wavelength cross-correlated variability for this flare we 
selected Segment 1 to correspond to MJD 55060 -- 55082. On visual inspection, we noticed that there is no 
significant activity in optical fractional polarization and radio fluxes, so, for our examination of 
multi-wavelength cross-correlated 
variability we only considered $\gamma-$ray, X-ray, optical and NIR fluxes. The correlated variability 
results are plotted in the top left panels of Figure \ref{fig:zdcf} and in Table \ref{tab:lagResults}. The results 
for this segment show simultaneous variation between the X-ray and $\gamma$-ray, and between the optical-V and NIR-J bands 
with peaks at zero lag.  During this period the ZDCFs suggest a lag of $\sim 3$ days between the X-ray/$\gamma$-ray and the
V/J bands with V/J leading.

\subsubsection{Segment 2 (MJD 55082 -- 55110)}    

From Fig. 1, we also notice that there are nearly simultaneous $\gamma-$ray, optical/NIR flares starting at 
$\sim$ MJD 55082 and peaking at MJD 55092. This flare has poor temporal coverage in X-rays,
though there is some evidence for an essentially simultaneous flare.  There is only modest variability  in 
optical polarization 
(which could be considered to show three small flares during that period, with one at the multi-band peak
and no evidence for changes in any of the radio fluxes. To search for multi-wavelength cross-correlated 
variability for this flare we defined Segment 2 as MJD 55082 -- 55110.  Again we only considered $\gamma-$ray, 
X-ray, optical and NIR fluxes in our analysis of multi-wavelength 
cross-correlated variability.  The
 results are plotted in the top right panel  of Figure \ref{fig:zdcf} and in Table \ref{tab:lagResults}.
The results for Segment 2 indicate simultaneous emission, with lags consistent with zero for all cross-correlations.

\subsubsection{Segment 3 (MJD 55150 -- 55200)}    

Our main focus in the manuscript is the strong peculiar flare (PF hereafter) noticed as peaking in the
$\gamma-$ray and X-ray band at $\sim$ MJD 55168 and declining to much lower flux states by $\sim$ MJD 55177 (see Fig. 1). 
To include the entire period of the activity in these bands we selected the Segment 3 to  span MJD 55150 -- 55200. 
On visual inspection, the  $\gamma$-ray, X-ray, optical, and NIR fluxes are strongly correlated 
during the flare period; however, these are anti-correlated with the PD during the decaying period of 
the flare between the vertical dashed lines in Fig.\ 1 and the PA goes through a large rotation during the flare.
Our DCF analyses for Segment 3  are plotted in the bottom panels of Figure \ref{fig:zdcf}; the left panels
give the multi-wavelength flux correlations as discussed above, whereas the right panels show fluxes against PA and PD. 
We did not use radio
flux data in the cross-correlation analysis as the only significant activity seen is a nearly monotonic rise in the 36.8 GHz flux 
density that appears to have begun before this flare period and to continue afterwards. The flux correlation 
results for this segment suggest simultaneous emission with lags consistent with zero for fluxes in 
$\gamma-$ray, X-ray and optical/NIR bands, at least once the gap in V and J coverage toward the end of Segment 3
is taken into account.   However, the PA/PD correlations with fluxes show significant anti-correlations
for PA against all fluxes  at zero lag, while the PD shows negligible correlations at zero lag with any of the  fluxes and 
marginal indications for  (see bottom right panels of Figure \ref{fig:zdcf}
and the bottom portion of Table \ref{tab:lagResults}. 

As can been seen from
the summary of previous work presented in the Introduction, the combination of such
correlations has never been reported for this, or any other, blazar. We note that
the optical/NIR flare starts earlier than do the X-ray and $\gamma$-ray flares, but
the peak fluxes in all these bands are essentially co-temporal.  The flare in X-ray
and $\gamma$-ray lasts for $\sim$10 days, which is the nearly same duration noticed
in earlier $\gamma$-ray flares (Bonning et al.\ 2009; Vercellone et al.\ 2009; Gaur et al.\ 2012).   
Unless there is a rather unlikely coincidence, in the sense that the strong X-ray
variation is independent of the other bands but just happened to coincide with them,
these combined LCs strongly suggest that the dominant  regions for optical through
$\gamma$-ray  production are co-spatial. The PA rotations are also strong during
the rising phase of the flare and during the post flare phase, indicating that the
region producing  the large flux changes possesses a strong, dominant magnetic field
direction, but one that is changing rapidly. 

Unfortunately, the radio observations were rather sparse during the strong flare,
but while our higher frequency radio LCs show  slow upward trends that start before
the flare and continue after it, there is no evidence for any rapid changes in these
radio bands coincident with the strong flare; furthermore, the lower frequency radio
LCs are consistent with constant fluxes throughout this period.  However, the 230
and 86 GHz light curves and 43 GHz VLBI measurements do show correlations (Jorstad et al.\ 2013).
This type of behavior is typical of blazar flares, where variations at cm-radio
wavelengths usually lag those at optical and mm-bands, as can be explained by standard
shock-in-jet models (e.g.\ Marscher \& Gear 1985; Hughes, Aller \& Aller 1985).

\begin{table}\label{tab:lagResutls}
  \centering
  \caption{Lag results for all the segments (in days)}
  \label{tab:lagResults}
  \begin{tabular}{cccc}
  \hline
Light curves    & 55060-55082             &  55082-55110   & 55150-55200 \\ 
		&Segment 1 &	Segment 2 & Segment 3 \\ \hline
$\gamma$ vs XRT & $-0.24^{+0.80}_{-0.45}$ & $-0.09^{+0.52}_{-1.80}$ & $-0.32^{+0.58}_{-0.54}$ \\ 
$\gamma$ vs V & $-3.9^{+2.1}_{-1.6}$      & $-0.2^{+0.4}_{-0.2}$    & $+0.3^{+4.2}_{-1.1}$ \\
$\gamma$ vs J & $-0.87^{+0.29}_{-2.11}$   & $-2.2^{+2.1}_{-1.1}$    & $+2.5^{+2.5}_{-1.9}$ \\
XRT vs V 	& $-5.6^{+2.4}_{-0.5}$    & $+0.08^{+0.41}_{-0.47}$ & $+10.1^{+1.9}_{-3.5}$ \\
XRT vs J 	& $-5.4^{+1.8}_{-0.6}$    & $-0.38^{+0.78}_{-0.71}$ & $+0.8^{+7.5}_{-1.8}$ \\
V vs J 		& $-0.02^{+0.20}_{-0.20}$ & $-0.05^{+0.38}_{-0.28}$ & $-0.86^{+0.69}_{-0.14}$\\
\hline
\multicolumn{4}{c}{flux vs PA for Segment 3 (55150-55200)}\\
\hline
$\gamma$ vs PA 	& $-1.0^{+1.2}_{-3.2}$    &    &  \\ 
XRT vs PA 	& $-5.6^{+4.1}_{-1.1}$    &    &  \\
V vs PA 	& $+0.1^{+1.5}_{-0.2}$    &    &  \\
J vs PA 	& $-8.3^{+8.5}_{-0.8}$    &    &  \\
\hline
\end{tabular}
 \end{table}

\subsection{Spectral Energy Distributions}\label{subsec:SEDs}

\begin{figure*}
\includegraphics[scale=0.65]{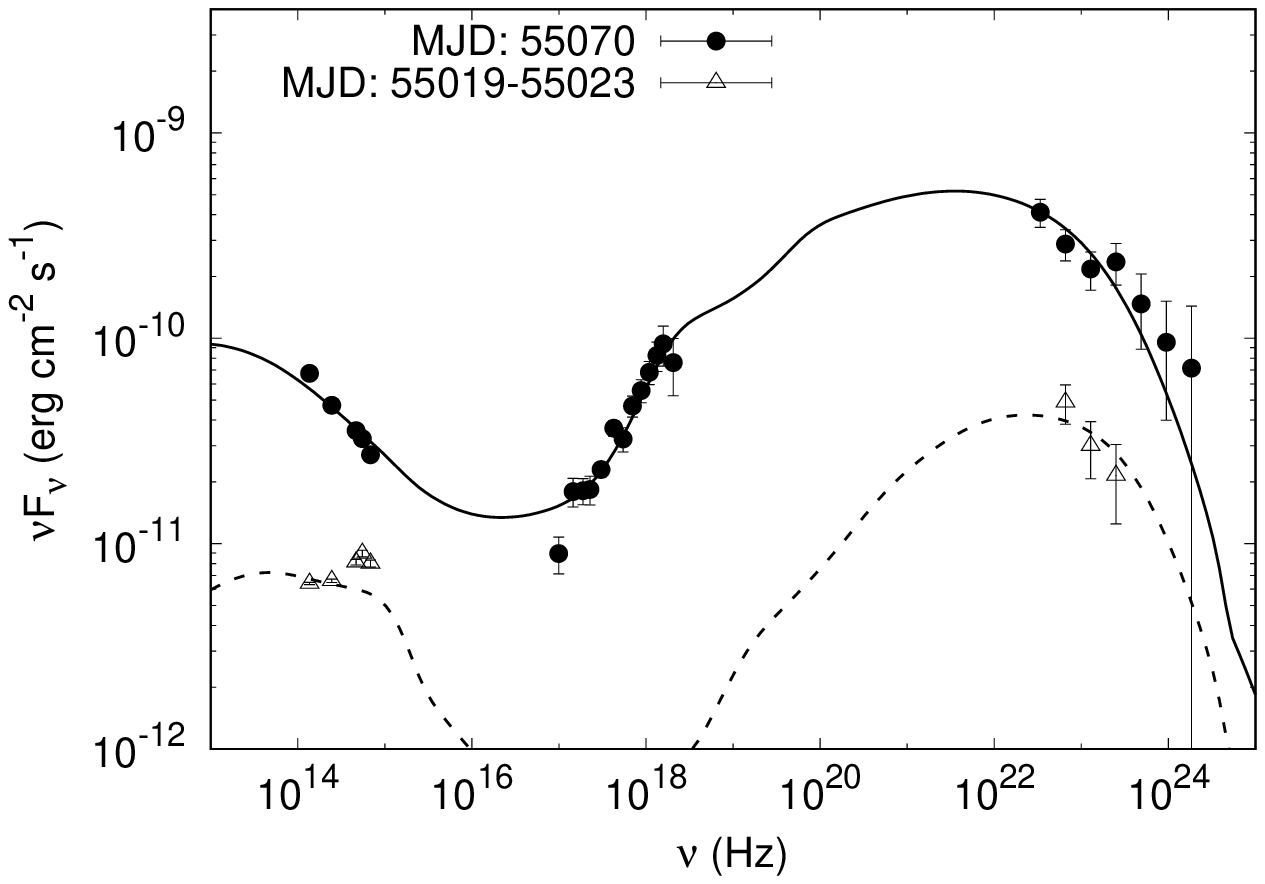}
\includegraphics[scale=0.65]{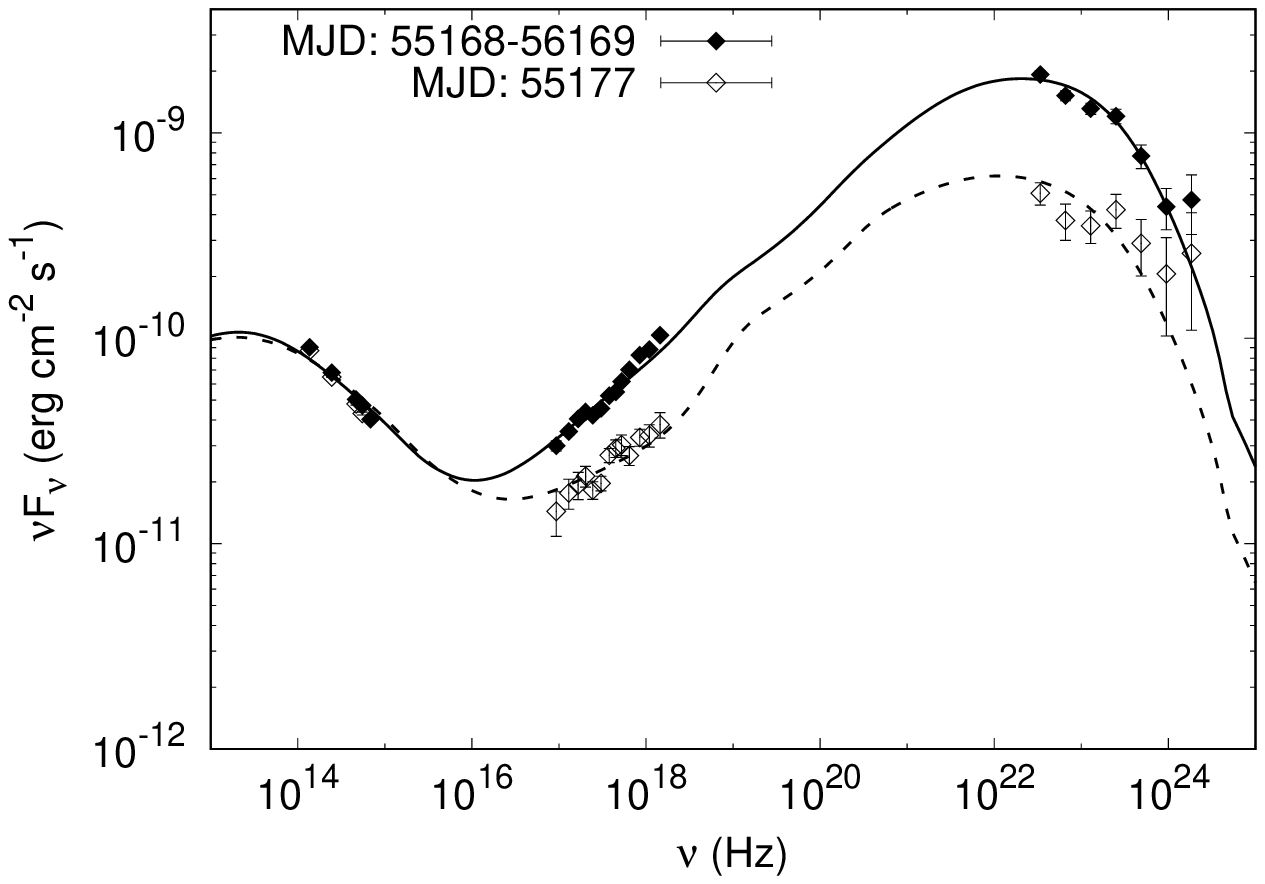}
\caption{Multi-wavelength SEDs during: (left) a quiescent period  from MJD 55019-55023
where it is lowest in all  bands for which we have data and at the peak of the first
flare; (right) and around the two epochs during the PF marked with dashed lines in
Segment 3 of Fig.\ 1, corresponding to the peak and during the decline of
that largest flare (see \S\ref{subsec:SEDs} for details).}
\label{fig:mwSEDs}
\end{figure*}

Figure \ref{fig:mwSEDs} shows SEDs at four epochs extracted from Fig.\ 1. The
first is for a quiescent state (MJD: 55019-55023), corresponding to the lowest
brightnesses in optical-NIR along with a low $\gamma$-ray flux, although there were
no X-ray data available then. The second is at the first  significant rise in all
bands (MJD: 55070). The third epoch, from MJD 55168--55169, is at the peak of the
peculiar flare (PF) for all of the $\gamma$-ray -- NIR bands. For the final selected epoch,
the SED is extracted from MJD 55177, during the decaying phase of the PF when the high-energy
bands have nearly reverted to their pre-flare levels, whereas the optical and NIR
emissions have barely begun to decay. 

We model all the SEDs assuming a one zone model with synchrotron and IC processes
arising from a smoothed broken power-law particle spectrum with indices derived
from the optical and X-ray data whenever available, following the approach of Kushwaha et al.\ (2013). 
Those bands can be attributed to emissions from single components (synchrotron and
SSC, respectively) and thus directly reflect the particle spectrum. In contrast, the
$\gamma$-ray spectrum appears to require inclusion of contributions from additional
components: IC of the broad line region (BLR) photons and IC of IR photons from
the torus around the central engine. During both these epochs, the flux changes
significantly within a day in some bands, giving a reasonable constraint to the
size of $\sim$1 lt-day ($\sim \rm 5\times 10^{16}$ cm). The resulting model fluxes
are the curves plotted in Fig.\ 3  and the values of the model parameters are given
in Table \ref{tab:parSED} (see \S\ref{sec:Discussion} for more details).

\begin{table}
\centering
\caption{\bf SED parameters}
\begin{tabular}{lcccc}
\hline
&&Epoch &&\\
&&(MJD)&&\\\hline
Parameter & 55019-23& 55070  &  55168-9 & 55177  \\
\hline
particle index before break$^\dagger$   & 2.0  & 2.6  & 2.1 & 2.3 \\
particle index after break$^\ddagger$   & 4.1  & 4.2  & 4.0 & 4.0   \\
magnetic field (Gauss)                  & 1.9  & 3.22 & 1.5 & 2.7  \\
equipartition fraction$^{\ast}$         & 0.05 & 0.2  & 1   & 0.9  \\ 
Doppler factor                          & 12   & 17   & 16  & 16  \\
particle break energy$^{\ast\ast}$       & 1114 & 594  & 706 & 662 \\
logarithm of jet power                  & & & &   \\
 (erg s$^{-1}$)                         & 45.5 & 47.2 & 46.8& 46.5  \\
minimum particle energy$^{\ast\ast}$ & 30   & 8    & 15  & 20  \\
\hline
\multicolumn{5}{l}{size of emission region: $\rm 5 \times 10^{16}$ cm} \\
\multicolumn{5}{l}{maximum particle energy$^{\ast\ast}$: $5\times 10^{4}$} \\
\multicolumn{5}{l}{torus covering factor: 0.3;  IR-torus temperature: 1200 K} \\
\multicolumn{5}{l}{$^\dagger$from X-ray;  $^\ddagger$from optical-NIR (except for 55019-23)}\\
\multicolumn{5}{l}{$^\ast$ particle energy density/magnetic energy density} \\
\multicolumn{5}{l}{$^{\ast\ast}$in units of electron rest mass energy} \\
\end{tabular}
\label{tab:parSED}
\end{table}

\subsection{Black Hole Mass Estimation}

\begin{figure*}
\centering
\includegraphics[width=170mm]{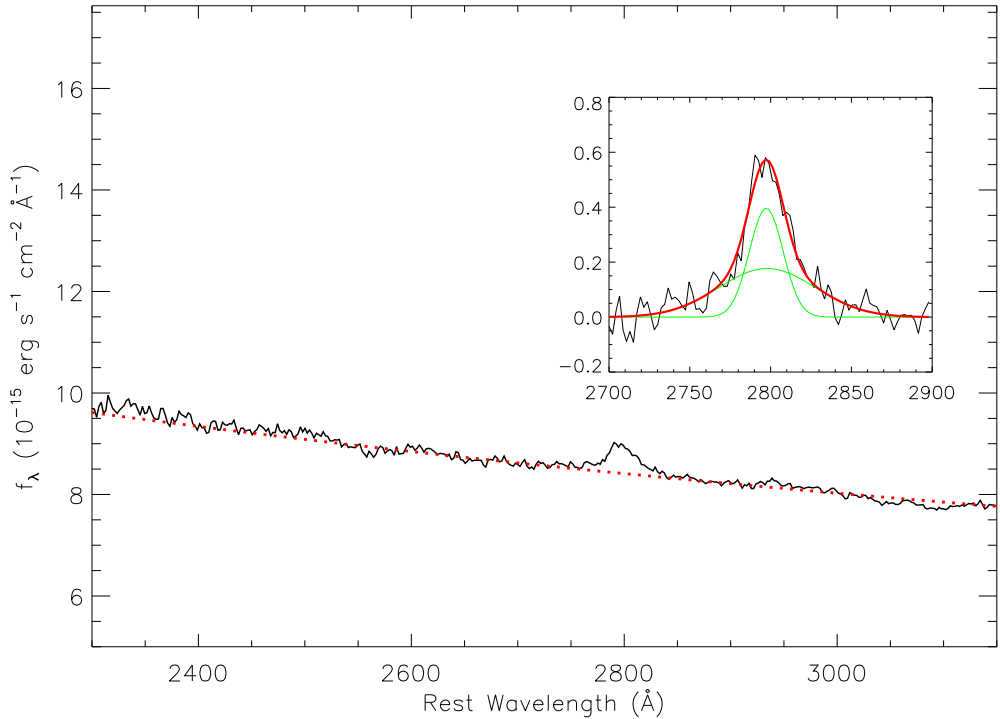}
\vspace*{-0.5in}
\caption{Steward observatory spectrum of 3C 454.3. The continuum modeled with a
single power-law is plotted as a smooth dotted line. The residual emission
line spectra after subtracting the power-law continuum is shown in the inset for
the Mg II region with the lower two smooth curves (green), giving the modelled broad
line components and the upper smooth curve (red), the entire modelled line profile.}  
\label{fig:spec}
\end{figure*}

Using different black hole (BH) mass estimation methods and multi-wavelength data,
the mass of the super-massive BH of the FSRQ 3C 454.3 previously has been estimated
in the broad range of 0.5 -- 4.5 $\times$ 10$^{9}$ M$_{\odot}$ (Gu et al.\ 2001; Woo
\& Urry 2002; Liu et al.\ 2006; Sbarrato et al.\ 2012).  As we have focused on the
multi-wavelength flaring event during December 3 -- 12, 2009 we downloaded a Steward
observatory archive optical spectrum of this blazar taken nearly simultaneously (on
December 15, 2009). 
We analysed this spectrum following the procedure given in Guo \& Gu
(2014) and  the fitted spectrum is shown in Fig.\ 4. 

We model the continuum as a single power-law, with f$_{\lambda} \propto \lambda^{\alpha}$ and  found the spectral 
index to be $\alpha = -$0.680. 
Also present is the broad emission line from Mg II that can be well modeled by combination of two 
Gaussian with FWHM 3300 km s$^{-1}$. The BH mass is estimated using the broad Mg II line width and the continuum 
luminosity at 3000\AA ~($\lambda$L$_{3000} =$ 8.64 $\times$ 10$^{46}$ erg s$^{-1}$; see Vestergaard \& Osmer 2009). 
This gives us  M$_{BH} = 2.3 \ (\pm 0.5 ~{\rm dex}) \times 10^{9}$ M$_{\odot}$. Our BH mass estimate is thus in agreement with  previous results  given the substantial errors present in all these methods.

\section{Discussion and Conclusions}\label{sec:Discussion}

Several competing models have been used  to explain the earlier flare events seen in
multiwavelength observations of blazars. Here we will  mention only some of those that have 
considered optical polarization properties to one degree or another. 
In most of the previous observations of blazars including polarimetry (e.g., Marscher et
al.\ 2008; Sasada et al.\ 2010; Marscher et al.\ 2010; Jorstad et al.\ 2010), a smooth
rotation of the polarization angle with the rise in optical flux has been noticed on
long term polarimetric observations. This can be explained by a non-axisymmetric magnetic
field distribution or a curved trajectory of the dissipation/emission pattern (e.g., K{\"o}nigl \& Choudhuri 1985; 
Marscher et al.\ 2008).  The large swings of polarization can be explained by a ``swinging jet" 
model where the angle the jet makes with our line of sight varies (e.g., Gopal-Krishna \& Wiita 1992). 
However, in the simplest version of this model, involving only Doppler factor variations, the fluxes in all bands should 
change similarly, along with a swing in the  in PA.  That is  is not the case here:  modest changes 
in optical--IR fluxes are seen along with substantial increments in X-ray and $\gamma-$rays.
If variability arises from helical magnetic field structures, the  observed polarization can be 
calculated following Lyutikov et al.\ (2005) and Raiteri et al.\ (2013).

The degree and direction of visible light polarization changed drastically during the
giant 20-day $\gamma$-ray flare in 3C 279 observed  by Abdo et al.\ (2010). Their
observations  unambiguously connect the $\gamma$-ray and visible-light emission regions,
showing that they emerge from essentially the same location. They then argue that
the changing polarization properties are indicative of motion in the jet: as a blob
of gas flows around a bend in the jet, for example, the changing angle between the
direction of the blob's motion and our line of sight can reproduce essentially the
observed changes in the angle and degree of polarization. However, no explicit
modeling was carried out to quantify this scenario.

Marscher et al.\ (2008)  very nicely modeled the variations they observed in BL Lac 
during 2005--2006 by considering a shock traveling along a spiral path, and we now
summarize their picture.  It assumes that activity originating close to the black
hole inserts energy into a portion of the jet's area. This would be seen as a knot
of emission as it propagates through both acceleration and collimation zones and
the Doppler beaming of synchrotron emission grows as the knot accelerates along
its spiral path  (Marscher et al.\ 2008). In this model the knot's synchrotron
output increases, producing  the bulk of the emission from BL Lac from the
optical through $\gamma$-ray bands until the disturbance
exits the zone of helical path. The greatest beaming, and hence the strongest flare,
in the LC occurs during the final spiral, when the Lorentz factor of the jet is
very high and the velocity of the knot is closest to our line of sight. When the
flare dominates the optical flux, we see the optical polarization vector rotate.
Including projection effects and relativistic aberration in the model allows the
optical PA data to be well fit and the similar PA seen at 7mm VLBA measurements
supports this picture. Owing to synchrotron self absorption the first flare is
not seen in the radio LCs.  A second flare occurs when the knot crosses the radio
core, which is  identified with a standing shock by Marscher et al.\ (2008). 

The relative constancy of the PA during these observations of 3C 454.3 in 2009,
aside from during and immediately after the major flare, implies that the 
above scenario was not being observed in this FSRQ at this time. 
However, recently Larinonov et al.\ (2013) have extended the Marscher et al.\ (2008) model to 
examine multi-wavelength variations
of a major flare in the blazar  S5 0716$+$714. 
They note that minor flares often both precede and follow major optical outbursts. 
They interpret this in terms of oscillating Doppler beaming of the
emission  (e.g.\ Camenzind \& Krockenberger
1992).  In the Larinonov et al.\ (2013) picture, the series of flares occur
 when our viewing angle to the propagating shock wave is smallest. 
Many different flux and polarization
behaviors could  be reproduced if the large number of parameters of the model
are varied appropriately (Larinonov et al.\ 2013).

Recently, Mohan \& Mangalam (2015) presented a fully relativistic (including GR effects) model of variability 
in AGN by considering  synchrotron emitting blobs in helical motion along a funnel or cone 
shaped magnetic surface anchored to the accretion disk near the black hole. The simulated 
light curves for the Mohan \& Mangalam (2015) model include light bending, time delay and 
aberration as well as both Doppler and gravitational shifts. They find that the beamed intensity 
has a systematic phase shift with respect to that from the simpler  special relativistic model 
of Camenzind \& Krockenberger (1992). The results indicate that a realistic magnetic surface 
geometry in a general relativistic framework is needed to describe how orbital features in the 
jet change the observed emission, at least in the vicinity of the black hole. For these unique 
observations of 3C 454.3, the critical results we need to explain are the systematic changes in 
the PA and in the polarization fraction during the modest change in optical-IR flux. A model along 
the lines of Mohan \& Mangalam (2015) takes advantage of GR effects when the source is close to 
the black hole and so bends, as invoked, e.g., by Abdo et al.\ (2010), can  explain the optical 
PA variation.  But it also exploits the helical path to explain the PD variation (e.g. Marscher 
et al. 2008; Larianov et al.\ 2013) and so has a better chance of explaining this complicated 
behavior.

The the overall boost factor $g$ is given by (see Mohan \& Mangalam 2015)
\begin{equation}
g = \frac{E_{\mathrm{obs}}}{E_{\mathrm{em}}} = (1-2 M/R)^{1/2} D = \frac{(1-2 M/R)^{1/2}}{\Gamma (1-\beta \cos \xi)}
\label{gfacdef}
\end{equation}
where $M$ is the black hole mass, $R$ is the distance (in mass units), $D$ is the Doppler factor, 
$\Gamma$ the bulk Lorentz factor of the blob, while $\xi$ is the angle between the direction of 
the photon to the observer and the instantaneous velocity of the blob. The fact that the optical 
flux  does not seem to change much indicates the near constancy of $\xi$ for the given viewing 
angle. For an appropriate choice of model parameters, as observed from GR simulations in a conical
geometry, the light curves can flatten out (see Fig.\ 11 in Mohan \& Mangalam 2015).

We can take the blob  to be equivalent to a mini-jet having a constant rest frame emission and 
polarization properties that is following a bent helical path.
The observed degree of polarization for  synchrotron emission coming from the region of
helical magnetic fields is found using $P=P_{\rm max} \sin^2 \chi^{\prime}$,
where $\chi'$ is the viewing angle in the jet rest frame.  This angle and  the 
observed viewing angle, $\chi$ are related through the Lorentz transformation 
\begin{equation}
\sin \chi'= {{\sin \chi} \over {\Gamma (1-\beta \cos \chi)}}
\end{equation}
where, as usual, $\Gamma$ is the bulk Lorentz factor of the plasma. Note that while the built-in GR features 
of this fully relativistic model are particularly useful
for time keeping, with respect to the clock of a distant observer, of the trajectory of blobs starting from 
near the black hole, the kinematic effects themselves are neglible when $r \gg r_g$.

As discussed in Lyutikov \& Kravchenko (2017), 
the plane of polarization rotation
lies in the projection of plane formed by line of sight and the velocity field
on the sky and so the PA rotation can  result from changes in magnetic field
topology or Doppler boosting or a combination of these. On the other hand,
the lag of $\gamma$-ray emission behind the NIR-optical output, which has been observed in this source
during other episodes \citep[e.g.][]{2017MNRAS.464.2046K},  can result from a steeper decline of 
the external radiation field as compared to the magnetic field, as argued by \citet{2012ApJ...754..114H}
and shown by \citet{2012ApJ...760..129J}. Further, a zero lag between X-rays and
$\gamma$-rays, but both lagging with respect to the NIR-optical strongly suggests
that the X-ray emission has a substantial contribution from the process dominating
the $\gamma$-ray band. Thus, a time dependent model is required to actually
understand the relative effect for both these situations. However, SED modeling can
still be used to understand the observation by exploring the relevant quantities
expected to cause such variation. In the SEDs displayed in Fig \ref{fig:mwSEDs}
those corresponding to the peak and decay portions of the PF show essentially no change in optical-IR
fluxes while X-rays and $\gamma$-rays both show substantial changes. At the same time,
the PA rotates by $180^{\circ}$. The SED for the first flare, on the other hand,
shows a substantial hardening at X-rays with a relatively steeper $\gamma$-ray spectrum
compared to the peculiar flare, while the quiescent SED provides some evidence for the
presence of a thermal bump in the optical-UV and displays
 a steeper $\gamma$-ray spectrum compared to the others.

In FSRQs, within the framework of a one-zone leptonic origin of the synchrotron
emission, the explanation of $\gamma$-ray spectra require external Comptonization
and can have contributions from multiple photon fields such as the BLR and/or torus,
depending on the location of the emission region along the jet. Due to contributions
from multiple components, the $\gamma$-rays can exhibit different spectra while the
IR-optical region may have an essentially constant spectral slope. 
Further, as the IR-optical component is synchrotron emission, the observed
spectrum is directly related to the relativistic particle spectra and in this case
the IR--optical (except when a thermal (accretion disc) bump contaminates this regime 
(as may be the case here for the quiescent SED) and X-ray spectra can be used to obtain the particle indices.
With this, the relative contributions of external IC from the BLR and IR regions can be determined by
reproducing the $\gamma$-ray emission, if the IR temperature is known.

We wish
to gain insights into the physics using a minimal number of variations in the parameters that can give
rise to  SEDs that are  consistent with understanding the PA rotations during
the PF and the observed lag during the PF, with the optical-NIR changes leading the X-ray/$\gamma$-ray ones.
Hence we fixed the IR to a 1200 K black body with a covering fraction of 0.3 as inferred
for the FSRQ PKS 1222$+$216 (Malmrose et al.\ 2011).
Parameters able to reproduce the observed SEDs are given in Table \ref{tab:parSED} and the
model fluxes for them are as shown in Fig.\ \ref{fig:mwSEDs}. This SED modeling 
alone can only constrain the magnitude of the magnetic field and
not its direction (hence, not the PA).  However, the only changes needed to explain
the essentially constant fluxes in the optical/IR and the very different fluxes in
the X-ray and $\gamma$-ray parts of the SED involve $B$, the particle normalization
(which can be derived from the equipartition factor given in Table 2)
and the particle spectrum before the break. So if there is no change to the Doppler
factor then the change in the strength and orientation of $B$ is required to occur
in such a way that it does not affect the overall emission in the low energy portion
of the SED (e.g. Joshi et al.\ 2016) and can also give rise to a hardening of the
particle spectrum with a Doppler to bulk Lorentz factor ratio of $\sim$ 1.6.  
On the other hand, the SED of the first flare is dominated by the magnetic field,
with X-rays resulting from EC, in contrast to the PF where the X-rays
are mainly from SSC. For the quiescent SED, the NIR-optical portion is
modelled as containing both synchrotron and thermal disc emission associated with
the mass of SMBH derived here ($\sim 2.3\times10^{10}$~ M$_\odot$). However,
most of the physical parameters are uncertain for the quiescent SED, in that substantially less
data was available then.  Furthermore, it should be noted that the modelled  thermal
bump from an accretion disk is an upper limit so that its signature is not visible
in the other three flare SEDs.

An improved model for synchrotron polarization in blazars, involving three-dimensional 
radiation transfer and assuming a standard shock-in-jet explanation for the flare in a 
jet with an originally dominant helical magnetic field, recently has  been developed 
(Zhang et al.\ 2014, 2015).  These simulations can reproduce the range of polarization
behaviors seen during earlier flares without requiring either bent or helical jet trajectories
(e.g. see Chandra et al.\ 2015). 

From approximately MJD 55080 to 55165, the PA generally stays at the same value, while the PD  rises and falls.
These changes in PD are coupled with some smaller flares in multiple bands as well. These changes are  likely to 
arise from shock compression and acceleration. For example, Laing's (1980) model argues that an increase in PD 
is due to the shock compression
of a turbulent field (cf.\ Marscher 2014); however, PA variations associated with such a field can be erratic. In our data,
especially the portion from MJD 55120 to 55130, where the PA has nearly no variation while the PD increases from 
$\sim5$\% to $\sim 15$\%, it appears that the PA variation is not as  erratic as expected from a turbulent field. 
So we suggest that the background field indeed is likely to be generally dominated by the toroidal component of a
helical magnetic field (Zhang et al.\ 2014, 2015). This is also suggested in the study by Jorstad et al.\ (2013) where it 
was observed that the quiescent state of 3C 454.3 during the current observation period was associated with the alignment of 
the optical polarization PA with the jet opening angle and was interpreted in terms of a well ordered toroidal magnetic field.
When a shock compresses the toroidal component, the PD will
increase while the PA generally stays the same and the radiation of this entire period may be from the same
emission region.
From MJD 55165 to 55185, the PA completes a 180$^{\circ}$ rotation, accompanied by a strong variation in PD.
This is probably due to some significant change in the emission region and it is very reasonable to suppose that
reconnection may happen inside the emission region.  In that case the toroidal field component is dissipated and the reconnection strongly accelerates particles, leading to strong flares extending up to the X-ray and $\gamma$-ray bands.
Meanwhile, the poloidal component becomes dominant and triggers a PA rotation and PD variation.
Afterward, the emission region recovers to its initial magnetic topology.

In AGN jets the electric vector position angles (EVPAs) on parsec scales tend to 
have polarization orthogonal to the
jet in radio-loud quasars, while BL Lac objects usually evince polarization along the jet 
(e.g.\ Lyutikov et al.\ 2005). The basic shock mechanism produces 
fields compressed in the shock plane and so for transverse shocks it naturally yields  EVPAs along the jet.  
However, as shocks are normally intrinsically transient events, it is difficult to see how the jet could retain its 
polarization orientation over extensive distances. In addition, since internal shocks in relativistic jets 
normally are oblique (e.g.\ Hughes 2005), the bimodal distribution of jet EVPAs between the AGN classes is 
unexpected. Hence, the EVPAs in BL Lacs seem to be 
in disagreement with the basic shock model. An alternative interpretation of the jet polarization, which we favor, 
is that 
the flow carries large-scale helical magnetic fields. The polarization properties of a relativistic jet carrying 
helical magnetic fields can both reproduce the average properties of the jet polarization as well as the bimodal
distribution of the observed EVPAs (Mangalam, in preparation).

By our combining $\gamma$-ray, X-ray, optical, NIR and radio monitoring of the FSRQ 3C 454.3 in $\sim$ MJD 55000 
through MJD 55200 with optical polarimetric data we have found a flare with apparently unique characteristics.  
The flare is essentially simultaneous from $\gamma$-ray down to NIR energies, which is relatively uncommon, and 
the polarization behavior is also unusual. 
Clearly, additional simultaneous multi-band observational campaigns addressed at both FSRQs and BL Lacs that also 
include polarization measurements are necessary for a better understanding of the location and physical mechanisms 
behind the variety of variations in blazars.

\section*{ACKNOWLEDGMENTS}

We thank the anonymous referee for useful comments which helped us to improve the 
manuscript. We thank Prof. M. B\"{o}ttcher for discussions. ACG is partially supported 
by Chinese Academy of Sciences (CAS) President's International Fellowship Initiative (PIFI) 
(grant no.\ 2016VMB073). HG is sponsored by CAS Visiting Fellowship for Researchers from Developing 
Countries, CAS PIFI (grant no.\ 2014FFJB0005), supported by the NSFC Research Fund for International 
Young Scientists (grant no.\ 11450110398) and supported by a Special Financial Grant from the 
China Postdoctoral Science Foundation (grant no.\ 2016T90393). PK is supported by grants from the 
Brazilian agency FAPESP 2015/13933-0. This work of LCH was supported by National Key Program
for Science and Technology Research and Development grant 2016YFA0400702. The work of RB, ES, 
and AS  is partially supported by Scientific Research Fund of the Bulgarian Ministry of Education 
and Sciences under grant DO 02-137 (BIn-13/09). MFG is supported by the National Science Foundation 
of China (grants 11473054 and U1531245) and by the Science and Technology Commission of Shanghai 
Municipality (grant 14ZR1447100). UMRAO was supported by Fermi GI grants NNX09AU16G, NNX10AP16G, 
NNX11AO13G, NNX13AP18G and by NSF grant AST-0607523. 

This paper has made use of up-to-date SMARTS optical/near-infrared light curves that are available 
at www.astro.yale.edu/smarts/glast/home.php. SMARTS observations of Large Area Telescope-monitored 
blazars are supported by Yale University and Fermi GI grant NNX 12AP15G, and the SMARTS 1.3-m 
observing queue received support from NSF grant AST-0707627. Data from the Steward Observatory 
spectropolarimetric monitoring project were used. This program is supported by Fermi Guest 
Investigator grants NNX08AW56G, NNX09AU10G, NNX12AO93G, and NNX15AU81G. This 
work made use of data supplied by the UK Swift Science Data Centre at the University of Leicester.
This research has made use 
of the NASA/IPAC Extragalactic Database (NED) which is operated by Jet Propulsion Laboratory, California 
Institute of Technology, under contract with the National Aeronautics and Space Administration.

\clearpage

\end{document}